\documentclass[a4paper,english]{article}
\usepackage[T1]{fontenc}
\usepackage[utf8]{inputenc}
\usepackage{units}
\usepackage{amsmath}
\usepackage{amssymb}
\usepackage{esint}



\usepackage{babel}
\begin{document}

\title{What is the meaning of non-uniqueness of FRW and Schwarzschild metrics?}

\author{Leonid V. Verozub, \\
Kharkov National University\\
lverozub@gmail.com}

\maketitle

\begin{abstract}
It is shown that any theory of gravitation, based on the hypothesis
of the geodesic motion of test particles must be invariant under geodesic
(projecive) mappings of the used space-time. The reason is that due
to invariance of the equations of geodesic lines under a continuous
group of transformations of the coefficients of affine connection,
there is a wide class of transformations of the geometrical objects
of Riemannian space-time which leaves invariant the equations of motion
of test particles. The FRW metric in cosmology and the Schwarzschild
metric are a good example to make sure that the standard space-time
metrics does not determine the gravitational field unequivocally.

\end{abstract}

\section{Introduction}

The equations of motion of test bodies play a fundamental role in
the classical field theory. They give evidence for the existence of
the field and allow us to find its properties. The functions appearing
in these equations are characteristics of the field.

In the case of gravity, the gravitational equations of motion of test
particles are invariant under some group of transformations of the
Christoffel symbols $\Gamma_{\beta\gamma}^{\alpha}$ -- of geodesic
transformations \cite{key-1}: 
\begin{equation}
\overline{\Gamma}(x)_{\beta\gamma}^{\alpha}=\Gamma(x)_{\beta\gamma}^{\alpha}+\psi_{\beta}(x)\delta_{\gamma}^{\alpha}+\psi_{\gamma}(x)\delta_{\beta}^{\alpha},\label{eq:GammaTransform}
\end{equation}
 where $\psi_{\alpha}$(x) is an arbitrary gradient covector field.
It is easiest to see it, if the coordinate $t=x^{0}/c$ is used as
a parameter in geodesic line:

\begin{equation}
\ddot{x^{\alpha}}+(\Gamma_{\beta\gamma}^{\alpha}-c^{-1}\Gamma_{\beta\gamma}^{0}\dot{x}^{\alpha})\dot{x}^{\beta}\dot{x}^{\gamma}=0.\label{eq:geodetics}
\end{equation}

It seems obvious that the Christoffel symbols $\overline{\Gamma}_{\beta\gamma}^{\alpha}(x)$
and $\Gamma_{\beta\gamma}^{\alpha}(x)$ describe the same physical
gravitational field, just as the 4-potentials in classical electrodynamics,
connected by a gauge transformation $A_{\beta}\rightarrow A_{\beta}+\partial_{\beta}\phi(x)$.

Such a gauge transformation of the Christoffel symbols induces a corresponding
transformation of the metric tensor $g_{\mu\nu}$, the curvature tensor
$R_{\alpha\beta\gamma}^{\delta}$, and the Ricci tensor $R_{\alpha\beta}$.
So, all these objects in themselves have no more physical meaning
than 4-potentials $A_{\mu}$ in electrodynamics. Evidently due to
this fact the equations of gravitational field, based on the hypothesis
on free motion on geodesic lines, must be invariant under the geodesic
(projective) transformations. It is well known that even vacuum Einstein's
equations do not satisfy this condition \cite{key-2}%
\footnote{We mean the original Einstein's equations, and not frequently considered
its mathematical generalization%
}.

In Riemannian space-time, eqs.(\ref{eq:GammaTransform}) are equivalent
to the mappings of the space-time $V$ with the metric tensor $g(x)_{\alpha\beta}$
to the space-time $\overline{V}$ with the metric tensor $\overline{g}(x)_{\alpha\beta}$
defined by the following PDE:

\begin{equation}
\overline{g}(x)_{\alpha\beta;\gamma}=2\psi(x)_{\gamma}\overline{g}_{\alpha\beta}(x)+\psi(x)_{\alpha}\overline{g}_{\gamma\beta}(x)+\psi(x)_{\beta}\overline{g}_{\alpha\gamma}(x),
\end{equation}
 where a semicolon denotes a covariant derivative with respect to
$x^{\alpha}$ in $V$. There exists extensive literature on the investigation
of the possibility of geodesic mapping $V\rightarrow\overline{V}$
based on these equations \cite{key-3}.

Consequently, every solution of Einstein's equations in any coordinate
system gives in general only one of many physically equivalent metrics
\cite{key-4} .

\section{FRW metric}

It is recently this fact has been discovered independently in \cite{key-20}
for the case of the FRW metric. It was noted in this paper (and after
that in \cite{key-8}) that the line element of this cosmological
model admits one-parameter transformations of the metric tensor that
leaves unchanged non parameterized geodesics.

But the truth is that this is not a random fact. This is not a specifics
of the FRW metric. The truth is that the metric, Christoffel symbols,
or the curvature tensor define the gravity field only up to geodesic
transformations, which should play the role of gauge transformations
in any geometrical theory of gravitation\cite{key-4}.

Due to simplicity of the FRW metric, consideration of the consequences
of such geodesic equivalence of metrics is especially simple.

Consider the line elements of a Riemannian space-time $V$:

\begin{equation}
ds^{2}=b(t)\, dt^{2}+a(t)\,\sigma_{ik}(x^{1},x^{2},x^{3})\, dx^{1}dx^{k}.\label{eq:ds}
\end{equation}
 It is known \cite{key-3-1} that geodesics of such metric are the
same as the ones of the space-time $\overline{V}$ with the line element
\begin{equation}
\overline{ds}^{2}=B(t)\, dt^{2}+A(t)\,\sigma_{ik}(x^{1},x^{2},x^{3})\, dx^{k}dx^{k}.\label{eq:dsprime}
\end{equation}

where

\begin{equation}
B(t)=\frac{b(t)}{[1+q\, a(t)]^{2}},\label{eq:A}
\end{equation}
 
\begin{equation}
A(t)=\frac{a(t)}{1+q\, a(t)},\label{eq:B}
\end{equation}
 and $q$ is an arbitrary constant .

Consider briefly the proof of this important fact.

Contracting (2) with respect to $\alpha$ and $\beta$, we obtain
$\overline{\Gamma}_{\beta\gamma}^{\beta}=\Gamma_{\beta\gamma}^{\beta}+(n+1)\,\psi_{\beta}.$
Consequently, 
\begin{equation}
\psi_{\beta}=\frac{1}{2(n+1)}\,\frac{\partial}{\partial x^{\beta}}\ln\left|\cfrac{\det\overline{g}}{\det g}\right|,\label{eq:psi}
\end{equation}
 which shows that in the case under consideration only $0$-component
of $\psi_{\alpha}$is other than zero.

The useful for us components of the Christoffel symbols of $V$ are:

\[
\overline{\Gamma}_{00}^{0}=b'(t)/2\, b,\quad\overline{\Gamma}_{11}^{0}=a'(t)/2b(t),\quad\overline{\Gamma}_{10}^{1}=a'(t)/2\: a(t),
\]
 and the same components of $\overline{V}$: 
\begin{equation}
\overline{\Gamma}_{00}^{0}=B'(t)/2\, B,\quad\overline{\Gamma}_{11}^{0}=-A'(t)/2B(t),\quad\overline{\Gamma}_{10}^{1}=A'(t)/2\: A(t).
\end{equation}
 Then eqs. (\ref{eq:GammaTransform}) gives the following equations

\begin{equation}
\frac{A'(t)}{A(t)}-\frac{a'(t)}{a(t)}=2\,\psi_{0},\label{eq:dif1}
\end{equation}

\begin{equation}
\frac{B'(t)}{B(t)}-\frac{b'(t)}{b(t)}=4\,\psi_{0},\label{eq:dif2}
\end{equation}

\begin{equation}
\frac{A'(t)}{B(t)}-\frac{a'(t)}{b(t)}=0.\label{eq:dif3}
\end{equation}
 Therefore, $A/a=\exp(2\int\psi(t)_{0}dt)$, $B/b=\exp(4\int\psi(t)_{0}dt)$
where the integration constants are equal to $1$ because at $\psi(t)_{0}=0$
the functions $A(t)=a(t)$ and $B(t)=b(t)$. Consequently, $B(t)/b(t)=(A(t)/a(t))^{2},$and
with (\ref{eq:dif3}) we obtain the differential equations 
\begin{equation}
A'(x)-A(t)^{2}\frac{a'(t)}{a(t)}=0,
\end{equation}
 which gives (\ref{eq:B}). Now from previous equation we obtain the
function $B(t)$ in the form (\ref{eq:A}).

On the contrary, if in eq. (\ref{eq:GammaTransform}) to set $\psi_{i}=0$
for i=1,2,3, and $\psi_{0}=-\nicefrac{1}{2\,}\partial$$\ln(1+qb(t))/\partial t$,
then eqs. \eqref{eq:dif1}, \eqref{eq:dif2}, and \eqref{eq:dif3}
are satisfied. Thus, with this choice of the covector field $\psi(x)_{\alpha}$,
the line element \eqref{eq:ds} at $b=-1$ is equivalent to \eqref{eq:dsprime}.
In other words, the both line elements have the same equations of
motion of test particles.

\section{Schwarzschild metric}

As another example, we show here that a static centrally symmetric
metric 

\begin{eqnarray}
ds^{2} & = & b(r)dr^{2}+r^{2}(d\theta^{2}+\sin^{2}\theta d\phi^{2})-a(r)dt^{2},\label{eq:ds-1}
\end{eqnarray}
(in particular, Shvartsshild metric) is not unique. Namely, in a given
coordinate system it has common geodesic lines with a metric of the
form 

\begin{equation}
\overline{ds}^{2}=B(r)\, dr^{2}+F(r)^{2}(d\theta^{2}+\sin^{2}\theta\, d\phi^{2})-A(r)\, dt^{2},\label{eq:dsprime-1}
\end{equation}
 where $A(x),B(x)$ and $F(x)$ are functions of r, depending on a
continuous parameter.

The Christoffel symbols for (\ref{eq:ds-1}) is given by

\begin{gather}
\Gamma_{rr}^{r}=\frac{1}{2}\frac{b'(r)}{b(r)},\;\Gamma_{r\theta}^{\theta}=\frac{1}{r},\;\Gamma_{r\phi}^{\phi}=\frac{1}{r},\;\Gamma_{rt}^{t}=\frac{1}{2}\frac{a'(r)}{a(r)},\;\Gamma_{\theta\theta}^{r}=-\frac{r}{b(r)},\label{eq:GammaV1}
\end{gather}
 
\begin{equation}
\Gamma_{\theta\phi}^{\phi}=\frac{\cos\theta}{\sin\theta},\;\Gamma_{\phi\phi}^{r}=-\frac{r\sin^{2}\theta}{b(r)},\;\Gamma_{\phi\phi}^{\theta}=-\sin\theta\cos\theta,\;\Gamma_{tt}^{r}=\frac{1}{2}\frac{a'(r)}{b(r)}.\label{eq:GammaV2}
\end{equation}
 The Christoffel symbols for \ref{eq:dsprime-1} are:

\begin{gather}
\overline{\Gamma}_{rr}^{r}=\frac{1}{2}\frac{B'(r)}{B(r)},\;\overline{\Gamma}_{r\theta}^{\theta}=\frac{F'(r)}{F(r)},\;\overline{\Gamma}_{r\phi}^{\phi}=\frac{F'(r)}{F(r)},\;\overline{\Gamma}_{rt}^{t}=\frac{1}{2}\frac{A'(r)}{A(r)},\;\overline{\Gamma}_{\theta\theta}^{r}=-\frac{F(r)\, F'(r)}{B(r)},\label{eq:GammaVprime1}
\end{gather}
 
\begin{equation}
\overline{\Gamma}_{\theta\phi}^{\phi}=\frac{\cos\theta}{\sin\theta},\;\overline{\Gamma}_{\phi\phi}^{r}=-\frac{F(r)\, F'(r)\sin^{2}\theta}{B(r)},\;\overline{\Gamma}_{\phi\phi}^{\theta}=-\sin\theta\cos\theta,\;\overline{\Gamma}_{tt}^{r}=\frac{1}{2}\frac{A'(r)}{B(r)},\label{eq:GammaVprime2}
\end{equation}
where a prime here and later denotes a derivative with respect to
$r$.

In view of this, Levi-chevita equations (\ref{eq:GammaTransform})
yields:
\begin{equation}
\frac{B'(r)}{B(r)}-\frac{b'(r)}{b(r)}=4\psi_{r}(r),\; F(r)F'(r)b(r)-rB(r)=0,\; A'(r)b(r)-a'(r)B(r)=0,\label{eq:eqs1}
\end{equation}

\begin{equation}
\frac{F'(r)}{F(r)}-\frac{1}{r}=\psi_{r}(r),\;\frac{A'(r)}{A(r)}-\frac{a'(r)}{a(r)}=2\psi_{r}(r),\;\psi_{\theta}(r)=\psi_{\phi}(r)=\psi_{t}(r)=0.\label{eq:eqs2}
\end{equation}

According to (\ref{eq:psi}) the function $\psi_{r}(r)$ . can be
written as 
\[
\psi_{r}=\partial\ln\chi/\partial r,
\]
where 
\[
\chi(r)=\left(\frac{\bar{g}}{g}\right)^{1/2(n+1)}.
\]
 Consequently,

\begin{center}
$\begin{array}{ccc}
B=b\chi^{4}; & A=a\chi^{2}; & F=\chi;\\
A'=a'\chi^{4}; & \left(F^{2}\right)^{'}=2r\chi^{4};
\end{array}$
\par\end{center}

Formulas for the function $F(r)$ are compatible only if the functions
$\chi(r)$ are the solution of the differential equations

\[
r\chi'(r)+\chi(r)-\chi(r)^{3}=0
\]
 which yields 
\[
\chi(r)=(1+kr^{2})^{1/2},
\]
where $k$ is an arbitrary constant. 

As a result, formulas which express the $A(r),B(r)$, and $F(r)$
by $a(r)$ and $b(r)$ are given by 

\begin{equation}
A(r)=\frac{a(r)}{1+kr^{2}},\; B(r)=\frac{b(r)}{\left(1+kr^{2}\right)^{2}},F(r)=\frac{r}{(1+kr^{2})^{1/2}},\label{eq:AandB}
\end{equation}

where $k$ is a constant satisfying apropriate physical conditions.

\section{Discussion}

It is obvious that the geodesics (projective) mappings of Riemannian
spaces should be considered as gauge transformations of the differential
equation, which is used to determine the geometrical characteristics
of gravity in any theory based on Einstein's hypothesis of the motion
of test bodies along geodesics of Riemannian space.

The fact that the connection coefficients and the metric tensor are
determined up to an arbitrary geodesic mapping, does not mean that
our physical space-time has a projective symmetry. No doubt the physical
space-time is locally pseudo-Euclidean and the notion of length has
a physical sense. This fact means that the geometric characteristics
of the physical space-time (the coefficient of the connection or metric
tensor) can not be directly identified with the characteristics of
the gravitational field, they are not observable variables of the
field. Such variables must be geodesically invariant.

However, the possibility to define observable variables of gravitational
field exists. For example, although Christoffel symbols are not not
be viewed as the observable characteristics of gravity, there are
symbols of Thomas, which are geodesically invariant objects. They
are not tensors. However, in the presence of a flat background metric,
a tensor object from the symbols of Thomas can be formed \cite{key-4}.

As for metrics, there are two possibilities to compare such theory
with observations.

Firstly, we can use solution of the field equations at some selected
gauge condition, just as we do it with solutions of the Einstein equations
at a selected coordinate conditions.%
\footnote{From a fundamental point of view, the problem of observables in general
relativity has not been solved.%
} (It is used in \cite{key-4}).

Secondly, there is an object that is geodesically invariant generalization
of the metric tensor $g_{\alpha\beta}$ if we consider the metric
tensor as the 4-components of some 5-dimensional tensor in the spirit
of a 5-dimensional interpretation of geodetic maps dating back to
Thomas \cite{key-10} and Veblen \cite{key-12}.

Let $X^{A}(A=0\div4)$ be homogeneous coordinates of points in the
tangent space of the space-time manifold with an arbitrary factor,
which is conveniently labeled as $\exp(x^{4})$. Then, in addition
to coordinate transformations 
\begin{equation}
\overline{x}^{\alpha}=\overline{x}^{\alpha}(x^{0},x^{1},x^{2},x^{3})\label{eq:4-transformations}
\end{equation}
 we must also take into account the change of this factor by transformation
of the fifth coordinate
\begin{equation}
\overline{x}^{4}=x^{4}+\log\rho\label{eq:5-transformation}
\end{equation}
 where $\rho$ is an arbitrary function of $x^{\alpha}$. In this
auxiliary five-dimensional manifold, we can define the geometric objects,
which are transformed through (\ref{eq:4-transformations}) and (\ref{eq:5-transformation}).
In particular, a tensor transforms as follows

\[
\overline{Q}_{AB}=Q_{CD}\frac{\partial x^{C}}{\partial x^{A}}\frac{\partial x^{D}}{\partial x^{B}},
\]
 where capital letters range from 0 to 4.

The equation
\[
G_{AB}X^{A}X^{B}=0
\]
 defines the quadric for which the equation of light cone $g_{\alpha\beta}dx^{\alpha}dx^{\beta}=0$
is an asymptotic. Tensor $G_{AB}$ determines a metric. In this case,
$g_{AB}=G_{AB}/G_{44}$ is a projective tensor, such that $g_{44}=1.$
If we define $f_{A}=g_{A0}$, it follows from the transformation law
of $G_{AB}$, that $f_{A}$ is a covariant projective vector which
transforms under pure 4-transformation of coordinates $ $as 
\[
\overline{f}_{4}=f_{4};\overline{\; f}_{\alpha}=f_{\beta}\frac{\partial x^{\beta}}{\partial\overline{x}^{\alpha}},
\]
 and under pure transformation of 5-coordinate (i.e under projective
transformations) as

\begin{eqnarray*}
\overline{f}_{4} & = & f_{4},\;\overline{f}_{\alpha}=f_{\alpha}-\frac{\partial\log\varrho}{\partial x^{\alpha}}.
\end{eqnarray*}

The tensor $g_{AB}$ is of the form 
\begin{equation}
g_{AB}=\left(\begin{array}{cc}
1 & f_{\alpha}\\
f_{\beta} & g_{\alpha\beta}
\end{array}\right)
\end{equation}
 where $g_{\alpha\beta}$ is an affine tensor which can be identified
with the metric tensor of space-time. It follows from the transformation
law (\ref{eq:4-transformations}) and (\ref{eq:5-transformation})
that the object
\begin{equation}
\overline{g}_{\alpha\beta}=g_{\alpha\beta}-f_{\alpha}f_{\beta}\label{eq:gobservaible}
\end{equation}
 is invariant under the transformation of the fifth coordinate, and
hence is invariant under the geodesic (projective) maps of space-time.

What is a field of $f_{\alpha}(x)$ ? It follows from the transformation
properties of the vector $f_{\alpha}$ with respect to the transformation
of 5-th coordinate that it can be a kind of a gradient invariant physical
field similar to the 4-potential of electromagnetic field which was
used in the Kaluza-Klein model. However, this vector can also be formed
from the components of the metric tensor, since the Christoffel symbols
$\Gamma_{\beta}=\Gamma_{\alpha\beta}^{\alpha}$ have the same transformation
law under the transformation of 5-th coordinate. This possibility
was used in paper \cite{key-11}.

From this point of view Einstein's equation is very similar to some
gauge-invariant equations that describe gravity in a fixed gauge.
The simplest equation of this kind, which do not contradict the available
observations, are proposed in \cite{key-4}.


\begin{thebibliography}{References}
\bibitem[1]{key-1} Eisenhart L., Riemannian Geometry, Princeton Univ.
(1926)

\bibitem[2]{key-2} Petrov A., Einstein Spaces , New-York-London,
Pergamon Press (1969).

\bibitem[3]{key-3-1} Mikes J., Journ. of Math. Sci., 78, 311 (1996)

\bibitem[4]{key-3} Mike$\check{s}$ J., Kiosak V., Van$\check{s}$urov$\acute{a}$
A. Geodesic mappings, Olomouc, (2008)

\bibitem[5]{key-4} Verozub L., Ann. der Phys., 17, 28, (2008)

\bibitem[6]{key-20} Neurowsky P., arXiv:1003.1503v1, (2010)

\bibitem[7]{key-8}Gibbons G, and Warnick C., arXiv: 1003.3845




\bibitem[9]{key-10}Thomas T. Y., Nat. Acad. Sci., USA 11, 199 (1925)

\bibitem[12]{key-12} Veblen O., Quart. Journ. Math. 1, 66 (1930)

\bibitem[11]{key-11} Verozub L. and Kochetov A., Grav. and Cosmol.
6, 246 (2000) \end{thebibliography}
\end{document}